\documentstyle[aas2pp4]{article}

\lefthead{TAKAHASHI et al.}
\righthead{[Ne {\small II}] 12.8 MICRON IMAGES OF FOUR GALACTIC ULTRACOMPACT H {\small II} REGIONS}

\begin{document}

\sloppy

\title{{[Ne {\normalsize\bf II}]} 12.8 MICRON IMAGES OF FOUR GALACTIC ULTRACOMPACT H {\normalsize\bf II} REGIONS : IONIZED NEON ABUNDANCE AS A TRACER OF THE IONIZING STARS}

\author{H. TAKAHASHI$^{1}$}
\affil{Department of Astrophysics, School of Science, Nagoya University, Furo-cho, Chikusa-ku, Nagoya, 464-8602, Japan; nori@u.phys.nagoya-u.ac.jp}

\and

\author{H. MATSUHARA, H. WATARAI and T. MATSUMOTO}
\affil{Institute of Space and Astronautical Science, 3-1-1 Yoshinodai, Sagamihara, Kanagawa
229-8510, Japan}

\altaffiltext{1}{Present affiliation : Research Assistant of Advanced Computational Science and Technology of Japan Science and Technology Corporation} 

\begin{abstract}

	We present the results of ground-based imaging spectroscopy of the {[Ne {\small II}]} 12.8 $\micron$ line emitted from the ultracompact (UC) H {\small II} regions; W51d, G45.12+0.13, G35.20$-$1.74 and Monoceros R2, with 2$\arcsec$ spatial resolution. We found that the overall distribution of the {[Ne {\small II}]} emission is generally in good agreement with the radio (5 or 15 GHz) VLA distribution for each source. The Ne$^{+}$ abundance ({[Ne$^{+}$/H$^{+}$]}) distributions are also derived from the {[Ne {\small II}]} and the radio maps. As for G45.12$+$0.13 and W51d, the Ne$^{+}$ abundance decreases steeply from the outer part of the map toward the radio peak. On the other hand, the Ne$^{+}$ abundance distributions of G35.20$-$1.74 and Mon R2 appear rather uniform. These results can be interpreted by the variation of ionizing structures of neon, which is primarily determined by the spectral type of the ionizing stars. We have evaluated the effective temperature of the ionizing star by comparing the Ne$^{+}$ abundance averaged over the whole observed region with that calculated by H {\small II} region models based on recent non-LTE stellar atmosphere models: 39,100 $^{+1100}_{-500}$ K (O7.5V$-$O8V) for W51d, 37,200 $^{+1000}_{-700}$ K (O8V$-$O8.5V) for G45.12$+$0.13, 35,000$-$37,600 $^{+1500}_{-600}$ K (O8V$-$O9V) for G35.20$-$1.74, and $\leq$ 34,000 K ($\leq$ B0V) for Mon R2. These effective temperatures are consistent with those inferred from the observed Ne$^{+}$ abundance distributions.

\end{abstract}



\section{INTRODUCTION}
\label{sec:intro}

	Ultracompact (UC) H {\small II} regions, which are known to be bright compact Galactic radio sources, are thought to be powered by massive young stars still embedded in their parent molecular clouds (Wood \& Churchwell 1989 ; hereafter WC). They are very small (r $\leq $ 0.1 pc) and dense (high electron number density : $n_{e} \geq$ 10$^{4}$ cm$^{-3}$), and have a large emission measure (EM $\geq $ 10$^{7}$ pc cm$^{-6}$). The spectral type of the ionizing star is uncertain for most of the UC H {\small II} regions because the ionizing stars are heavily extinguished by the dust clouds surrounding the UC H {\small II} regions, which prevent optical and even near-infrared observations even though the ionizing stars are very luminous. On the other hand, mid-infrared fine-structure lines do not suffer so heavily from extinction by the dust, and thus are good probes of the UC H {\small II} regions.

	Since the warm dust which surround the UC H {\small II} regions re-radiates nearly all of the absorbed stellar photons in the far-infrared, the UC H {\small II} regions are known as strong {\it Infrared Astronomical Satellite} ({\it IRAS\/}) point sources, with flux densities $\geq$ 10$^{4}$ Jy at 100 $\micron$. WC pointed out that the total FIR luminosity derived from the {\it IRAS\/} flux is larger than that expected from the spectral type of the ionizing star estimated from the radio continuum flux for most of the UC H {\small II} regions they observed. Two possible interpretations were presented by WC; 1) strong dust absorption of ionizing photons inside the UC H {\small II} region, and 2) presence of a stellar cluster including a large number of non-ionizing low-mass stars, which substantially contribute to the total luminosity. However, due to the uncertainty in the dust extinction inside the UC H {\small II} region, it is difficult to distinguish which effect is dominant.

	In this paper, we propose the {[Ne {\small II}]} line at 12.814 $\micron$ as a probe of the spectral type of the ionizing star. The ionization potential of neon (Ne$^{0} \rightarrow$ Ne$^{+}$ : 21.56 eV) is higher than that of hydrogen, and therefore, this line is a good tracer of early-type (especially B0 or late-O) stars which form the UC H {\small II} regions. Moreover, the {[Ne {\small II}]} flux is not sensitive to $n_{e}$ as long as $n_{e} \leq n_{crit}$ (= 3.6 $\times$ 10$^{5}$ cm$^{-3}$; the critical density of the line; Lacy, Beck, \& Geballe 1982), and are thus good probes of the ionic abundance, even for dense UC H {\small II} regions. In addition, since the line is within the atmospheric window, the {\it N}$-$band (8$-$13 $\micron$), it is accessible from ground-based telescopes.

	The sizes of UC H {\small II} regions are very small; the typical source size of $\sim$ 0.1 pc corresponds to 2$\arcsec$ on the sky if the source is located at 10 kpc away from us. Thus the {[Ne {\small II}]} line observations in the  previous work could not spatially resolve the source due to their large beams (7$-$30$\arcsec$). The 2$\arcsec$ spatial resolution of the present work enables us to obtain the spatially resolved {[Ne {\small II}]} images of the UC H {\small II} regions, which can be directly compared with the radio VLA images (WC; Kurtz, Churchwell \& Wood 1994).

	In the recent work by Watarai et al. (1998; hereafter WMTM), we presented {[Ne {\small II}]} observational results for two UC H {\small II} regions, G29.96$-$0.02 and W51d, using the same instrumentation, and estimated the effective temperature of the ionizing star from the Ne$^{+}$ abundance, by using H {\small II} region models based on recent stellar atmosphere models. Here, we present new observational results for four UC H {\small II} regions including W51d, for which observations were performed over a wider area than that of WMTM. Because of the increased number of observed sources in this work, we can discuss the relation between the spectral type of the ionizing star and the Ne$^{+}$ distribution of the UC H {\small II} regions.

	The selected four UC H {\small II} regions are W51d, G45.12+0.13, G35.20$-$1.74, and Mon R2 from the WC source list. W51d, also known as W51 IRS 2, is the brightest region in the W51 star forming region. The total FIR luminosity is estimated to be 3 $\times$ 10$^{6}$ L$_{\odot}$ (Erickson \& Tokunaga 1980), which is a factor of 3 brighter than the bolometric luminosity of an O4 zero-age main sequence (ZAMS) star, whereas the Lyman continuum luminosity ($N_{Lyc}$) derived from the radio flux corresponds to that of an O5.5 star. G45.12+0.13 is a well-studied UC H {\small II} region at many wavelengths. Lumsden \& Puxley (1996) showed very good positional coincidence between the Br$\gamma$ peak and the radio continuum peak. WC noted that an O4 (single) star or O5 (cluster) star is required to explain the FIR luminosity (1.7 $\times$ 10$^{6}$ L$_{\sun}$), whereas a single O5 star can account for the Lyman continuum luminosity (log $N_{Lyc} \sim$ 49.53). G35.20$-$1.74 is part of the W48 complex. According to WC, the total FIR luminosity (2.8 $\times$ 10$^{5}$ L$_{\sun}$) measured by {\it IRAS\/} is larger than that expected from the spectral type (O7.5) derived from the radio observation. As for Mon R2, its FIR luminosity is dominated by IRS 1. Downes et al. (1975) estimated the spectral type of the ionizing star to be B0 ZAMS or earlier star based on the radio observation. Howard, Pipher, \& Forrest (1994) also reported that the ionizing star located at IRS 1$_{\rm SW}$ (see Figure 1 or Table 4 in their paper) is B0$-$B3. The Galactocentric distances (R$_{\rm G}$) of these sources are similar to that of the Sun ($R_{\rm G}$ = 6$-$9 kpc), and hence we assume a constant elemental abundance of neon equal to the local interstellar abundance (hereafter LIA) : 9.9 $\times$ 10$^{-5}$ (Simpson et al. 1998).

	This paper is organized as follows. In $\S$2, we describe the observations, and the data reduction. In $\S$3, we present the [Ne {\small II}] images, and derive the Ne$^{+}$ abundance map. In $\S$4, we discuss the effective temperature for each source by using calculations based on the latest stellar atmosphere models in relation to the ionization structure. We also give some remarks for individual sources. Finally, $\S$5 gives the summary of this work.

\section{OBSERVATIONS}

	The {[Ne {\small II}]} 12.8 $\micron$ line images were obtained at the Wyoming Infrared Observatory (WIRO) using a liquid helium cooled Mid-Infrared Fabry-Perot Imager (MIRFI; Watarai et al. 1996) developed at Nagoya University. MIRFI incorporates a Si:P 5 $\times$ 5 detector array sensitive to mid-infrared wavelength ($\leq$ 28 $\micron$), and a Fabry-Perot interferometer with spectral resolving power $\lambda/\Delta \lambda = 1600$ (190 km s$^{-1}$ in velocity). The plate scale with the WIRO 2.3-m telescope is 1$\/\farcs\/7$ pixel$^{-1}$ and so the field-of-view (FOV) is 10$\/\farcs\/$2 $\times$ 10$\/\farcs\/$2 including gaps between pixels.

	The observations of the UC H {\small II} regions were carried out on 4$-$11 October 1997. The telescope beam was mainly pointed at the radio peaks. However, since the FOV of MIRFI is not large enough to cover the whole H {\small II} region, we carried out a raster mapping. For each source, the raster mapping was carried out at various spatial positions around the radio continuum peaks. One frame of data was taken at each raster position. Each raster position was accurately determined by position analysis using observations of the guide stars. The spatial sampling interval of the raster frames is 0$\/\farcs\/3-0\/\farcs\/$8. After the flat-fielding using the blank sky data, the final images were made up by mosaicing of the images taken at all raster positions. Finally we could obtain wider spatial coverage images of each source. Details of observations are shown in Table 1 and 2. The secondary chopping frequency is 2.5 Hz, and the chopper throw was 40$\arcsec$ along the east-west direction. Furthermore, nearly all data were taken by nodding of the telescope beam by 40$\arcsec$ along the east-west direction in order to remove the residual sky emission features which were visible in the data with a single chopping beam.

\placetable{tbl-1}

	The {[Ne {\small II}]} spectral data were acquired by wavelength scanning of Fabry-Perot interferometer in ``LINE SCAN" mode (Watarai et al. 1996). In this mode, the interferometer scans around the {[Ne {\small II}]} line wavelength in 16 steps over the wavelength range of 0.034 $\micron$ (770 km s$^{-1}$ in velocity) with a step size of 0.0021 $\micron$ (48 km s$^{-1}$). It takes 12.8 sec to perform a LINE SCAN. We define 40 LINE SCANs, half of which are taken at each nodding position, as 1 {\it frame\/}, hence the total observing time of 1 frame is 512 sec. The 3$\sigma$ detection limit was estimated to be 6 $\times$ 10$^{-20}$ W cm$^{-2}$ pixel$^{-1}$ for 1 frame observing time.

\placetable{tbl-2}

	The absolute flux scale was calibrated using the continuum fluxes emitted from the mid-infrared bright stars $\alpha$ Ori, $\chi$ Cyg, and $\alpha$ Tau. The flux densities at 12 $\micron$ of these stars are assumed to be 2820 Jy, 850 Jy and 450 Jy, respectively. The flux of $\alpha$ Ori is taken from the {\it IRAS\/} Low Resolution Spectrometer (LRS) data (Olnon et al. 1986), the flux of $\chi$ Cyg is taken from Gillett, Low, \& Stein (1968), and the flux of $\alpha$ Tau is given in the mid-infrared spectra measured by Cohen \& Davies (1995). The uncertainty of the absolute flux calibration is estimated to be $\pm$ 12 \% including the uncertainty in the line-fitting procedure.

	The observed spectra at the [Ne {\small II}] peak positions are shown in Figure 1. Each spectrum has been fitted with a constant continuum and a Lorentzian line profile. The spectral resolution of MIRFI is larger than the intrinsic {[Ne {\small II}]} line width expected for UC H {\small II} regions ($\leq$ 100 km s$^{-1}$), and so the derived line profiles represent instrumental ones. The central wavelength and the wavelength scale were determined by the measurement of two nearby bright atmospheric emission features (12.832 $\micron$ and 12.870 $\micron$). As a result, the FWHM line width and the central wavelength were determined to an accuracy of $\Delta v$ $\leq$ 20 km s$^{-1}$ (3$\sigma$).

\placefigure{fig1}

	MIRFI is equipped with an {\it H}$-$band monitor for guiding the source into the MIR detector FOV and for checking the pointing accuracy. During each observation, bright SAO stars near the sources were pointed before and after every {[Ne {\small II}]} line observation of the sources. The resulting uncertainty in the position determination was estimated to be $\leq$ 1$\arcsec$. For the positional data of the guide stars, we used the latest {\it HST\/} guide star catalog.\footnote{ Taken from http://www-gsss.stsci.edu/support/rcat$_{-}$form.html}

\section{RESULTS}

\subsection{[Ne {\small II}] Images}

	As we described before, the raster mapping was carried out for each source. The line fluxes at same sky positions between the frames taken at different raster positions agree well within the statistical errors. Therefore, the relative flux calibration among all raster positions is reliable. The number of raster positions (= number of the frames) obtained for each source are listed in Table 2. Mon R2 was observed over a lot of frames, but only one-third of the radio emitting region was covered, since Mon R2 extends to more than 25$\arcsec$ wide. Furthermore, we took frames at only 5 raster positions for G35.20$-$1.74, hence the whole emitting region could not be covered.

	 The final {[Ne {\small II}]} images smoothed by a $1\/\farcs\/7$ FWHM (1 pixel of MIRFI detector array) Gaussian profile and corrected for the interstellar extinction are shown in Figure 2. The {[Ne {\small II}]} peak positions of all sources except for G45.12+0.13 show good agreement with the 5 or 15 GHz radio continuum peaks within the positional error. As for the correction for interstellar extinction, the following values are used; $A_{V}$ = 20 mag for W51d based on $J, H, K$, and radio observations by Goldader \& Wynn-Williams (1994), $A_{V} \sim$ 24 mag for G45.12+0.13 from the Br$\gamma$ imaging and the radio continuum observations by Lumsden \& Puxley (1996), $A_{V} \sim$ 23 mag for G35.20$-$1.74 derived from CO observations (Vall\'{e}e \& MacLeod 1990) and from the Br$\alpha$/Br$\gamma$ ratio (Woodward, Helfer, \& Pipher 1985), and $A_{V} \sim$ 23 mag for Mon R2, from the Br$\gamma$/Br$\alpha$ ratio with a relatively large beam (28$\arcsec$) by Natta et al. (1986). We assume a uniform extinction over the whole area of each source.

	However, the [Ne {\small II}] line is on the 10 $\micron$ silicate feature and the extinction law from NIR to MIR wavelength is still uncertain, especially toward UC H {\small II} regions. Moreover it may vary with location in the Galaxy.  $A_{V}$ also may have spatial distribution inside the sources. Hence the uncertainty due to extinction should be an additional $\simeq$ 20 \%.

\placefigure{fig2}

\subsection{Physical Properties and Ne$^{+}$ Abundance Maps}

	In order to obtain the Ne$^{+}$ abundance ({[Ne$^{+}$/H$^{+}$]}), we derive the physical properties using the observed {[Ne {\small II}]} flux and the 5 or 15 GHz radio continuum flux. The procedure is described in WMTM in more detail. First, we calculated the brightness temperature ($T_{b}$) and the optical depth ($\tau$) of each source from their radio continuum fluxes, and then the emission measure (EM) was derived using an expression of the free-free radiation including a correction for the optical depth effect. We estimated a dependence of the electron temperature to EM assuming Case-B recombination. The EM changes only a few percent under the physical condition of typical UC H {\small II} regions, even if $T_{e}$ changes from 5000 K to 15000 K. Accordingly we have considered the error of EM is mainly caused by systematic and absolute flux errors. Since the radio fluxes are quoted from the same WC survey data, the relative errors in the radio fluxes among the sources should be very small. The values of VLA image rms in a 1$\farcs$7 resolution element are only 1$-$4 \%. On the other hand, the absolute radio flux errors over the whole H {\small II} regions can be estimated from the differences between VLA observations and single dish observations, which are about 20 \% for all sources (e.g., Massi et al. 1985). Finally we have added this uncertainty for that of EM for ``TOTAL REGION". The {[Ne {\small II}]} emission line photons are emitted as a result of collisional excitation by electrons. For all sources observed, the critical density ($n_{crit}$ = 3.6 $\times$ 10$^{5}$ cm$^{-3}$ for the [Ne {\small II}] line) is much larger than the electron number density ($<$ 10$^{5}$ cm$^{-3}$) estimated from the EM derived from the radio flux for the angular scale of the MIRFI beam. Therefore we neglect the correction for the critical density ($n_{e}/n_{crit}\ \ll$ 1) and the line flux is assumed to be proportional to EM. Then the ionic abundance of singly ionized neon is expressed as:
\begin{equation}
\frac{\rm Ne^{+}}{\rm H^{+}} = \frac{F{\rm_{[Ne\ II]}\ (W\ cm^{-2})}}{2.3\times10^{-9}\ \Omega_{b}\ T_{e}^{-1/2}\ e^{-hc/\lambda k T_{e}}\ {\rm EM}}.
\end{equation}
Here $F_{\rm [Ne\ II]}$ is the intrinsic [Ne {\small II}] line flux corrected for interstellar extinction, $\Omega_{b}$ is the solid angle of the observed area, and $T_{e}$ is the electron temperature of the source.

	The observed and extinction-corrected {[Ne {\small II}]} flux, and derived physical properties are listed in Table 3. The ``TOTAL REGION" means the whole region in which the {[Ne {\small II}]} emission is appreciably (3$\sigma$) detected. The ``PEAK REGION" represents the 1$\/\farcs\/$7 $\times$ 1$\/\farcs\/$7 region (corresponding to 1 pixel of MIRFI) centered at each radio continuum peak.

	Additionally we assumed electron temperature for optically thin H {\small II} regions, however, the extreme peaks of W51d and G45.12+0.13 are thought to be not optically thin. WC estimated peak brightness temperatures of 9800 K and 8100 K with corresponding optical depths of 3.71 and 1.66 for W51d and G45.12+0.13, respectively (see their Table 17). They assumed T$_{e}$ = 10,000 K. Our assumptions of electron temperature for these two sources are 7700 K and 8000 K. These may not be suitable for the peaks due to the optical depth effect, because an electron temperature distribution does make a difference from flat distribution around radio peak as we assumed. When the integrated radio fluxes before smoothing over the 1.7 arcsec$^{2}$ beam at peaks compare with that of after smoothing, the differences are 2$-$12 \% for each source. These are the additional uncertainty in their radio flux, succeedingly T$_{b}$, $\tau$ and EM for ``PEAK REGION". On the other hand, averages for ``TOTAL REGION" may not be significantly affected by the high optical depth at extreme radio peaks.

\placetable{tbl-3}

	We also derived the Ne$^{+}$ abundance maps, using the {[Ne {\small II}]} images and the radio continuum images smoothed by the beam of MIRFI. These are shown in Figure 3. In order to aid the eye to see the trend of the Ne$^{+}$ abundance distribution, in Figure 4 we also show the cut views through the radio peak. The directions of the cut are shown in Figure 3 (A$-$A$^{\prime}$). As is seen in Figure 3, the Ne$^{+}$ abundance has its minimum at the radio peak for all sources, and increases steeply towards the outer regions, especially for W51d and G45.12+0.13 (the ratio of the Ne$^{+}$ abundance at the radio peak (1$\/\farcs\/$7 $\times$ 1$\/\farcs\/$7) to that of the total area ([Ne$^{+}$/H$^{+}]_{\rm peak}$/[Ne$^{+}$/H$^{+}]_{\rm total}$) are 0.35 $\pm$ 0.07 for W51d and 0.28 $\pm$ 0.04 for G45.12+0.13), which are ionized by relatively massive stars. On the other hand, the Ne$^{+}$ distributions are relatively uniform for G35.20$-$1.74 and Mon R2 ([Ne$^{+}$/H$^{+}]_{\rm peak}$/[Ne$^{+}$/H$^{+}]_{\rm total}$ are 0.87 $\pm$ 0.17 for G35.20$-$1.74 and 0.72 $\pm$ 0.12 for Mon R2), which are ionized by relatively low-mass stars. These results indicate that the Ne$^{+}$ distribution depends strongly on the spectral type of the ionizing star.

\placefigure{fig3}
\placefigure{fig4}

\subsection{Results for Individual Sources}

{\it a) W51d}

	WMTM reported 2.8 $\times$ 10$^{-18}$ W cm$^{-2}$ for the {[Ne {\small II}]} line flux observed in one raster position centered on the radio peak. The [Ne {\small II}] line flux (2.4 $\times$ 10$^{-18}$ W cm$^{-2}$ at the peak position) of the present work is in good agreement with that of WMTM within the flux calibration error. In this work this source was observed at many raster positions as mentioned in $\S$2, and hence we can derive the detailed spatial structure of the {[Ne {\small II}]} emission and that of the Ne$^{+}$ abundance. Figure 2$a$ shows that the extent of the {[Ne {\small II}]} emitting region is almost the same as that of the radio continuum emission. The observed {[Ne {\small II}]} line flux integrated over the whole area is 2.8 $\times$ 10$^{-17}$ W cm$^{-2}$, which is the highest among the sources observed. The cometary structure clearly visible on the radio image is not seen in the {[Ne {\small II}]} image, which may be due to the inferior spatial resolution of MIRFI. As shown by Figure 3$a$ and Figure 4$a$, the Ne$^{+}$ abundance at the radio peak is the lowest among the all sources.

{\it b) G45.12+0.13}

	The {[Ne {\small II}]} flux integrated over the whole observed area is 1.8 $\times$ 10$^{-17}$ W cm$^{-2}$. This value is lower than that obtained by the previous observations with relatively large beams; 3.1 $\times$ 10$^{-17}$ W cm$^{-2}$ in a 7$\/\farcs\/$5 beam by Herter et al. (1981), and 2.6 $\times$ 10$^{-17}$ W cm$^{-2}$ in a 10$\arcsec$ beam by Lester (1979). The reason for this discrepancy may be partly due to an insufficient coverage in the present work, since the {[Ne {\small II}]} emission probably extends to the outer region, especially to the south-west as shown in Figure 2$b$. So the total {[Ne {\small II}]} flux derived in this work is possibly a lower limit.

	Figure 2$b$ shows that the general outline of the {[Ne {\small II}]} emitting region is consistent with that of the radio continuum image, especially for the relatively low brightness part. Both the radio and the {[Ne {\small II}]} intensities generally decrease towards the north-east from the peaks in the south-west. Figure 3$b$ clearly shows that the Ne$^{+}$ abundance decreases steeply toward the radio peak, the same trend as that seen in W51d.

{\it c) G35.20$-$1.74}

	There have been many studies of this source as a part of W48 (e.g., Persi et al. 1997), however, the {[Ne {\small II}]} observation has not yet been reported. The extended emission in the {[Ne {\small II}]} image is in good agreement with the 15 GHz image by WC. Both images show an extended emission toward the south-west direction. This suggests the ionizing sources of both the H {\small II} region and the Ne$^{+}$ region are the same. The total observed {[Ne {\small II}]} line flux is 9.5 $\times$ 10$^{-18}$ W cm$^{-2}$ (20.9 $\times$ 10$^{-18}$ W cm$^{-2}$, after correction for interstellar extinction). The Ne$^{+}$ abundance distribution is uniform over the whole area. Furthermore the Ne$^{+}$ abundance at the peak is larger than that of W51d and G45.12+0.13.

{\it d) Mon R2}

	WC showed that the 5 GHz radio emission extends to more than $\sim$ 25$\arcsec$. The observed region by the [Ne {\small II}] line is a brightest part of the source (the south-west part; see Figure 2$d$). As shown in Figure 2$d$, the radio continuum and the {[Ne {\small II}]} emission distributions are in good agreement with each other. The {[Ne {\small II}]} flux integrated over the whole area is 2.0 $\times$ 10$^{-17}$ W cm$^{-2}$ consistent with the flux of 2.3 $\times$ 10$^{-17}$ W cm$^{-2}$ in the 7$\arcsec$ beam obtained by Herter et al. (1982). The Ne$^{+}$ abundance distribution is similar to that of G35.20$-$1.74.

\section{DISCUSSION}
\label{sec:discussion}

\subsection{Effective Temperature of the Ionizing Star}

\placefigure{fig5}

	The Ne$^{+}$ abundance distributions differ greatly between the sources as described in $\S$3.2. Here we estimate the effective temperature of the ionizing stars by comparing the observed Ne$^{+}$ abundance over the whole observed region with that given by H {\small II} region models based on the latest stellar atmosphere models. As treated by WMTM, we use the calculations of the line flux ratio by Stasi\'nska and Schaerer (1997), which are based on {\it CoStar} stellar atmosphere models (Schaerer et al. 1996a, 1996b, Schaerer \& de Koter 1997). In Figure  5 the calculated Ne$^{+}$ abundance is shown as a function of the stellar effective temperature taken from grid calculations with $n_{e} = 10^{4}$ cm$^{-3}$ and the LIA for elemental abundance of neon.\footnote{ The grid calculation adopted the solar abundance ($\rm [Ne/H] = 1.7 \times 10^{-5}$) as the total neon abundance ($\rm = [Ne^{total}$]). We assume the calculated $\rm [Ne^{+}/Ne^{total}]$ is valid even for the case of the local interstellar abundance. The electron temperature calculated in this model is very close to that observed : 7500$-$8000 K.} From the observed Ne$^{+}$ abundance for ``TOTAL REGION" (Table 3), we estimate the stellar effective temperatures of the sources to be as follows: 39,100 $^{+700}_{-500}$ K (corresponding to the spectral type of O7.5V$-$O8V ZAMS star) for W51d, 37,200 $^{+700}_{-700}$ K (O8V$-$O8.5V) for G45.12+0.13, 37,600 $^{+900}_{-600}$ K (O8V$-$O9V) for G35.20$-$1.74. As for Mon R2, an upper limit of 33,000 $^{+1000}$ K ($\leq$ B0V) is derived, because the Ne$^{+}$ abundance is as large as the LIA value of neon while the $CoStar$ model grids for $T_{\rm eff}$ $<$ 33,000 K are not available. Taking all errors into account, the upper limit becomes to be 34,000 K. The errors in the effective temperature are mostly due to the uncertainty in the absolute [Ne {\small II}] and radio fluxes. Even considering these errors, we can conclude that the effective temperature of the ionizing stars in W51d is the highest among the sources, and that of Mon R2 is the lowest. 

	Although we estimated the effective temperature assuming that total abundance of Ne is LIA, the existence of the Galactic gradient for abundances of various ions is reported (Simpson \& Rubin 1990, Henry \& Worthey 1999). The Galactic gradient in O/H, and almost certainly Ne/H as well, is $-$0.06 dex/kpc. The range in $R_{\rm G}$ of the observed H {\small II} regions is 3.1 kpc, which implies a difference in Ne/H of 0.18 dex or a factor of $\sim$1.5. Taking these into consideration, total Ne abundances for each source are [Ne/H] = 11.7 $\times$ 10$^{-5}$ for W51d, 11.1 $\times$ 10$^{-5}$ for G45.12+0.13, 12.5 $\times$ 10$^{-5}$ for G35.20$-$1.74, and 8.2 $\times$ 10$^{-5}$ for Mon R2. Using these values instead of LIA, the estimated stellar effective temperatures of the sources are as follows: 39,500 K (corresponding to the spectral type of O7.5V$-$O8V ZAMS star) for W51d, 37,500 K (O8.5V) for G45.12+0.13, 38,200 K ($\leq$ O8V) for G35.20$-$1.74, and $\leq$ 33,000 K ($\leq$ B0V) for Mon R2. 

\placefigure{fig6}

	The radial distribution of the Ne$^{+}$ abundance shown in Figures 3 and 4 can be qualitatively understood by a simple model of the ionization structure controlled solely by the stellar effective temperature. The model is described in Figure 6. In this simple model, the H {\small II} region is assumed to be spherically symmetric and the distribution of the electron number density is assumed to be uniform over the whole H {\small II} region. Since the ionization potential of Ne$^{+}$ (41.02 eV) is higher than that of Ne$^{0}$ (21.56 eV), the ratio of amount of Ne$^{+}$ ionizing photons to that of Ne$^{0}$ ionizing photons depends strongly on the stellar effective temperature. Thus the volume ratio of the Ne$^{++}$ region to the Ne$^{+}$ region decreases as the effective temperature decreases. The volume ratio of the Ne$^{++}$ sphere to the Ne$^{+}$ shell is determined so that the average [Ne$^{++}$/Ne$^{+}$] over the whole H {\small II} region is equal to that expected from the following effective temperature based on the model calculation shown in Figure 5: 39,000 K for Figure 6$a$, 37,000 K for 6$b$, 35,000 K for 6$c$, and 33,000 K for 6$d$, respectively. The average Ne$^{+}$ abundance distribution over the lines of sight crossing the equator of the sphere is also given in bottom of each panel of Figure 6. The observed Ne$^{+}$ distributions as cut-views shown in Figure 4 are well explained by the model; Figure 6$a$ corresponds well with the cut view of W51d, in the same way, 6$b$ to G45.12+0.13, 6$c$ to G35.20$-$1.74, and 6$d$ to Mon R2. However, clearly an actual H {\small II} region is not spherically symmetric, and thus the two-dimensional Ne$^{+}$ distribution shown in Figure 3 cannot quantitatively be modeled since the detailed structure of the actual H {\small II} region is unknown. We would emphasize, however, the general trend seen in the variety of the Ne$^{+}$ abundance distributions is consistent with this simple model.

	An alternative explanation for the cause of the relatively low Ne$^{+}$ abundance obtained for W51d and G45.12+0.13 may be that the [Ne {\small II}] emission is heavily extinguished by dust even though the line is in MIR wavelength. Especially there may be the localized variation in the extinction by dust. However, if the intrinsic [Ne {\small II}] fluxes from these sources are an order of magnitude higher that those presented in Table 3, then the extinction should be enormous: $A_{V} \sim$ 65 mag. Furthermore, clear near-infrared images of Br$\gamma$ line emission which are similar to VLA images have been reported for both sources (W51d: Goldader \& Wynn-Williams 1994; G45.12+0.13: Lumsden \& Puxley 1996). Hence it is not likely that such large amount of dust exists as a foreground screen.

	In Table 4, we compare the estimated spectral types with those derived from the FIR and the radio luminosity. Interestingly, for the luminous sources W51d and G45.12+0.13, the spectral types derived from the FIR and the radio luminosity are much earlier than that estimated by the present work, while for the less luminous sources G35.20$-$1.74 and Mon R2, both estimates are in good agreement. One possible explanation for this trend is that the ionizing source is not a single star but is a star cluster, especially for the luminous sources. The spectral type derived from the FIR luminosity may be earlier than the actual spectral type because in the relatively large FIR ({\it IRAS\/}) beam, numerous low-mass stars can contribute to the heating of the dust and hence increase the FIR luminosity, though they do not contribute to either the hydrogen or neon ionizing photons. The single-star spectral type derived from the radio luminosity (WC) may also be earlier than that of the actual one if the H {\small II} region is ionized by early type stars which do not emit neon ionizing photons.

	Moreover, the Ne$^{+}$ abundance would not be a good tracer of the spectral type of the principal ionizing stars of the H {\small II} regions formed by very luminous stars like W51d and G45.12+0.13, since Ne$^{+}$ is not the dominant species of neon for such sources. The observed {[Ne {\small II}]} image is likely to be contaminated by the contribution from less luminous ionizing stars, leading us to underestimate the spectral type of the principal ionizing star. We will further discuss this matter in $\S$4.2.2, especially for G45.12+0.13. In case of less luminous sources, the H {\small II} regions are likely to be ionized by a single star, and Ne$^{+}$ is the dominant species of neon. Therefore, the radio-derived spectral type agrees well with that derived in the present work.

\placetable{tbl-4}

	The {[Ne$^{++}$/Ne$^{+}$]} abundance ratio would be a better tracer of the spectral type of the ionizing stars, since this does not depend on the total elemental abundance of neon. However, as shown in Table 4, the spectral type derived by the MIR line ratio ([Ne {\small II}] 12.8 $\micron$ and [Ne {\small III}] 15.6 $\micron$; Simpson \& Rubin 1990) is much later than that derived by the radio luminosity. Since these MIR lines are observed with the large beam of {\it IRAS\/} LRS (25$-$30$\arcsec$), there might exist some contamination by less luminous ionizing stars. Observations of the MIR {[Ne {\small III}]} lines (15.6 $\micron$ and 36.0 $\micron$) at higher angular resolution can provide an answer to this problem.

\subsection{Remarks for Individual Sources}

{\it a) W51d}

	The derived effective temperature ($\sim$ 38,500 K) is similar to that concluded by WMTM. Figure 3$a$ and Figure 4$a$ indicate that the Ne$^{++}$ is the dominant species of neon in W51d.

{\it b) G45.12+0.13}

	The {[Ne {\small II}]} peak is not associated with the main radio peak (position A), but is associated with the second radio peak (position B; see Figure 2). Since the positional uncertainty of our observation is $\leq$ 1$\arcsec$, a spatial separation between two peaks of $\sim$ 2$\arcsec$ is significant. The Ne$^{+}$ abundance derived at the {[Ne {\small II}]} peak (position B) is twice as large as that of the radio peak (position A). One possible explanation for this excess of the Ne$^{+}$ abundance at position B is that there is a small H {\small II} region ionized by a relatively late type star at position B superposed on the major H {\small II} region ionized by a massive star located at position A. Crude estimates of the {[Ne {\small II}]} and radio fluxes originating from this small H {\small II} region are 2.1 $\times$ 10$^{-17}$ W cm$^{-2}$ and 0.87 Jy, respectively, giving the Ne$^{+}$ abundance in the small H {\small II} region as 8.6 $\times$ 10$^{-5}$, which indicates the existence of a B0 star ($T_{\rm eff}\sim$ 33,000 K : Figure 5). However there is no firm evidence for the existence of the small H {\small II} region, and it is noteworthy that the center velocities of the {[Ne {\small II}]} lines at both positions are the same within the error of the wavelength calibration ($\leq$ 20 km s$^{-1}$). Hence the small H {\small II} region, if it exists, should have almost the same proper motion as that of the major H {\small II} region. Alternatively, the excess of the Ne$^{+}$ abundance could be simply due to a geometrical effect. Since the effective temperature of the major ionizing star is relatively high, most of Ne$^{+}$ ions are ionized to a higher ionization state (Ne$^{++}$). Then we may expect that the Ne$^{+}$ region is distributed around the Ne$^{++}$ region like a thin shell, and position B may be simply the tangential component of the Ne$^{+}$ shell, enhancing the Ne$^{+}$ abundance.

	There are many measurements of Ne and other lines that can be used to model the H {\small II} regions and thereby determined the stellar types. Colgan et al. (1991) reported the results of observations of FIR lines. The derived effective temperature ($\sim$ 37,000 K) is nearly consistent with our result. However ratio of Ne$^{+}$/Ne$^{++}$ is inconsistent with our results. The difference is caused by as follows: Colgan et al. used LTE models (Kurucz 1979) to estimate the effective temperature (see Watarai et al. 1998). We re-estimated its using non-LTE ($CoStar$) models and the Ne$^{+}$/Ne$^{++}$ ratio they used, which became to be $\sim$ 34,000 K. Furthermore the beam size of their observations are larger than the size of sources, hence there may be some possibility of contaminations.

	In order to reveal the true picture, observations of both the {[Ne {\small II}]} line and {[Ne {\small III}]} line images at higher spatial resolution are essentially needed. The results will soon confirm by large aperture telescope observations like SOFIA ($Stratospheric Observatory For Infrared Astronomy$) and SUBARU.

{\it c) G35.20$-$1.74}

        From Figure 3$c$ and Figure 4$c$, the Ne$^{+}$ abundance is distributed uniformly like Mon R2 but is lower than the LIA of neon. One possible interpretation is that because the observed region does not cover the whole H {\small II} region, we may have selectively observed the ionized region in which Ne$^{++}$ is the dominant species of neon. As shown in Figure 4$c$, the Ne$^{+}$ abundance tends to increase gradually at the outer part. Furthermore, the Ne$^{+}$ abundance at the PEAK REGION is larger than those of W51d or G45.12+0.13 (Table 3), suggesting that the thickness of Ne$^{+}$ region toward the peak is relatively large, as shown in Figure 6$c$. Hence the Ne$^{+}$ abundance presented for the TOTAL REGION would be a lower limit of the Ne$^{+}$ abundance over the whole source and the corresponding stellar effective temperature (37,000 K) would be an upper limit. The cut view shown in Figure 4$c$ corresponds well with the average Ne$^{+}$ distribution presented in Figure 6$c$. These results indicate that there exists a slightly lower temperature star ($\sim$ 35,000 K) in G35.20$-$1.74, than those of W51d and G45.12+0.13, but higher than the one in Mon R2. Persi et al. (1997) proposed that these are B0$-$B3 stars. WC also reported that the spectral type of the ionizing star is O9$-$O9.5 based on the FIR flux and radio fluxes. Further observations with much wider spatial coverage than the present work are necessary to confirm this.

{\it d) Mon R2}

	The Ne$^{+}$ abundance map represents uniform distribution. The value of the Ne$^{+}$ abundance is larger than those of the other sources and is as large as the LIA of neon. This implies that the spectral type of ionizing star is not earlier than O9. Howard, Pipher, \& Forrest (1994) showed that IRS 1$_{\rm SW}$ is a B0$-$B3 star, consistent with our results. Furthermore, although with a larger beam (7$-$30$\arcsec$), Herter et al. (1982) reported that the low excitation lines ({[Ar {\small II}]} 6.99 $\micron$, {[S {\small III}]} 18.7 $\micron$ and {[Ne {\small II}]} 12.8 $\micron$) were detected significantly, while high excitation one ({[Ar {\small III}]} 8.99 $\micron$ and {[S {\small IV}]} 10.5 $\micron$) were not clearly detected, supporting that the spectral type is early B.

	We use $A_{V}$ = 23 mag as the correction for interstellar extinction (Natta et al. 1986). Natta et al. (1986) and Massi, Felli, \& Simon (1985) pointed out that different values of the extinction were derived from measurements with different beam sizes, implying that the extinction is not uniform. Howard, Pipher, \& Forrest (1994) presented the extinction map from Br$\alpha$ and Br$\gamma$ images, reporting an extinction $\sim$ 27.3 mag at the IRS 1$_{\rm SW}$. The Ne$^{+}$ abundance map presented here shows a slight increase toward the outside from the peak. Hence if the value $\sim$ 30 mag is used for the correction due to extinction at the peak, the Ne$^{+}$ abundance map would become more uniform.

\section{SUMMARY}

	Mid-Infrared {[Ne {\small II}]} line images of four Galactic UC H {\small II} regions with $\sim$ 2$\arcsec$ spatial resolution are presented. The main results and conclusions are as follows.

\begin{enumerate}
\item{The {[Ne {\small II}]} images are spatially resolved for all sources. In comparison with the radio VLA images, the {[Ne {\small II}]} line and radio continuum images are found to be generally in good agreement. We also derived the Ne$^{+}$ abundance maps. For W51d and G45.12+0.13, the Ne$^{+}$ abundance steeply decreases toward the radio peak and is lowest at the radio peak. This suggests the Ne$^{++}$ region dominates near the centers of these H {\small II} regions. On the other hand, the Ne$^{+}$ abundance is distributed rather uniformly for G35.20$-$1.74 and Mon R2, and the Ne$^{+}$ abundance is almost equal to the LIA of neon for Mon R2. These trends indicate that the spectral type of the ionizing star plays a major role in the Ne$^{+}$ abundance distribution.}

\item{We estimated the effective temperature of the ionizing stars from the Ne$^{+}$ abundance by using H {\small II} region models based on the latest stellar atmospheric models: 39,100 $^{+1100}_{-500}$ K for W51d, 37,200 $^{+1000}_{-700}$ K for G45.12+0.13, 37,600 $^{+1500}_{-600}$ K for G35.20$-$1.74, and $\leq$ 34,000 K for Mon R2. The range of the effective temperatures include the effect of Galactic gradient in Ne/H. We further proposed a simple model of the ionization structure of neon controlled solely by the stellar effective temperature. The averaged Ne$^{+}$ distribution derived from the present observations (Figure 4) and that from the models (lower graphs of each model in Figure 6) generally agree well. As a result, the effective temperature of the ionizing star for G35.20$-$1.74 is estimated to be $\sim$ 35,000 K.}

\item{We also presented several results and discussions for individual sources as follows; possible interpretations of the difference between [Ne {\small II}] and radio continuum peaks in G45.12+0.13, and the non-uniform effect of interstellar extinction for Mon R2.}

\end{enumerate}

	Finally, we have demonstrated in this paper that the character of the ionizing star in the UC H {\small II} region can be revealed by the spatially resolved MIR line images.

\acknowledgments

	We would like to thank C. Woodward, director of WIRO, J. Weger, and other WIRO staffs for their valuable assistance at WIRO. We would like to thank H. Shibai for useful advice and discussions. We wish to thank E. Churchwell for providing VLA observation data and D. Schaerer for providing calculated data of the {\it CoStar} models. We would thank C. P. Pearson and the referee for their valuable comments and suggestions which substantially improved the presentation of this work. Part of this work was done with the support of Grant-in-Aid for Scientific Research of Ministry of Education, Science, Sport, and Culture of Japan, No. 07044054, Research Fellowships of the Japan Society for the Promotion of Science for Young Scientists, and Japan Science and Technology Corporation.

\newpage

\figcaption[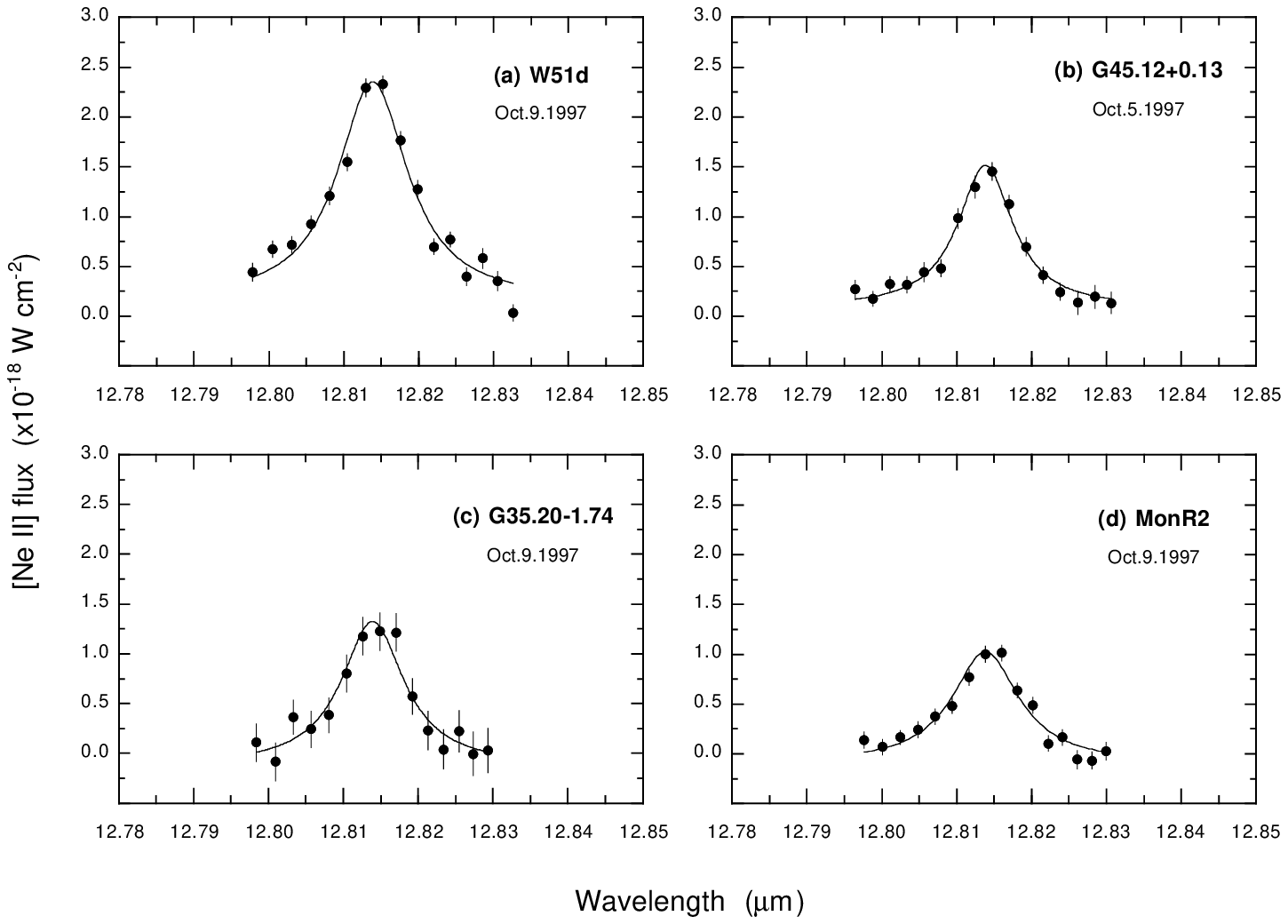]{Observed {[Ne {\small II}]} 12.814 $\micron$ emission spectra at the [Ne {\small II}] peaks of four UC H {\small II} regions. ($a$) W51d, ($b$) G45.12+0.13, ($c$) G35.20$-$1.74, and ($d$) Mon R2. These spectra have not yet been corrected for interstellar extinction. It should be noted that the intrinsic line fluxes are about twice as large, since $A_{V}$ is $\sim$ 20 mag for all sources (see text). These spectra are fitted with Lorentzian function (Fabry-Perot transmission profile) shown by solid line. Any continuum emission is not significantly observed. The error bars represent the 1$\sigma$ statistical error derived from the rms fluctuations over the observation of 40 LINE SCANs (512 sec in total). \label{fig1}}

\figcaption[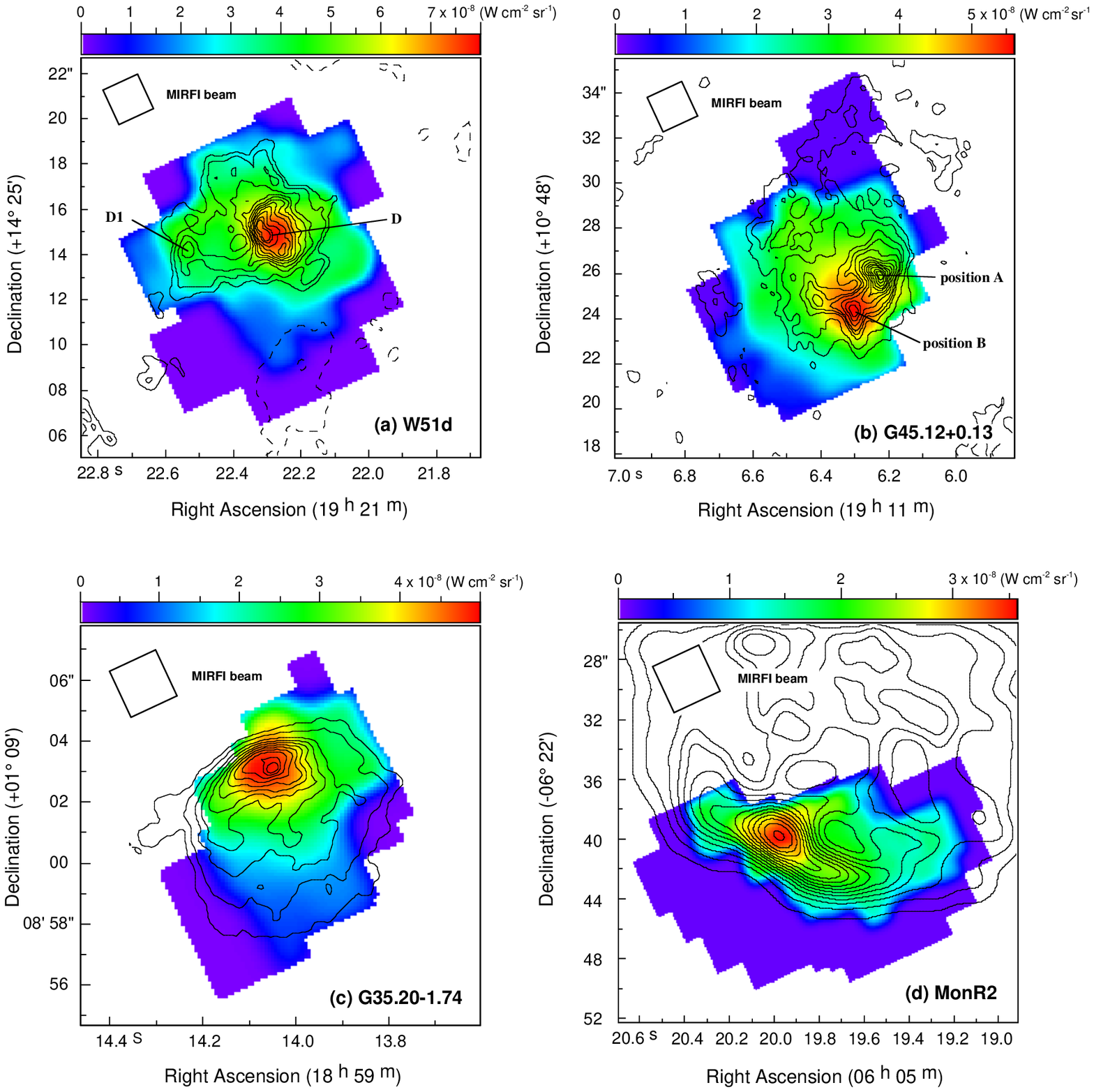]{False color images of the {[Ne {\small II}]} emission intensity corrected for the interstellar extinction: ($a$) W51d, ($b$) G45.12+0.13, ($c$) G35.20$-$1.74, and ($d$) Mon R2, superposed on the radio continuum images (contours). Radio data are taken from the 15 GHz observations of WC for all sources except for Mon R2, for which the 5 GHz observation of WC is taken. The peak flux density is 3.0033$\times$10$^{-1}$ Jy/beam and contoures are drawn at the levels $-$5.0, 5.0, 10, 20, 30, 40, 50, 60, 70, 80, 90, and 95\% of the peak for W51d; 2.4905$\times$10$^{-1}$ Jy/beam and 1.0, 3.0, 5.0, 10, 15, 20, 30, 40, 50, 60, 70, 80, 90, and 95\% for G45.12+0.13; 8.3436$\times$10$^{-2}$ Jy/beam and 5.0, 10, 20, 30, 40, 50, 60, 70, 80, 90, and 95\% for G35.20$-$1.74; and 8.6147$\times$10$^{-2}$ Jy/beam and 2.0, 9.0, 16, 23, 30, 37, 44, 51, 57, 64, 71, 78, 85, 92 and 99\% for Mon R2. The observed regions in the [Ne {\small II}] line are shown as the colored area. The spatial sampling intervals of the raster mappings are typically 0$\farcs$3 for W51d, 0$\farcs$7 for G45.12+0.13, 0$\farcs$8 for G35.20$-$1.74, and 0$\farcs$6 for Mon R2, respectively. \label{fig2}}

\figcaption[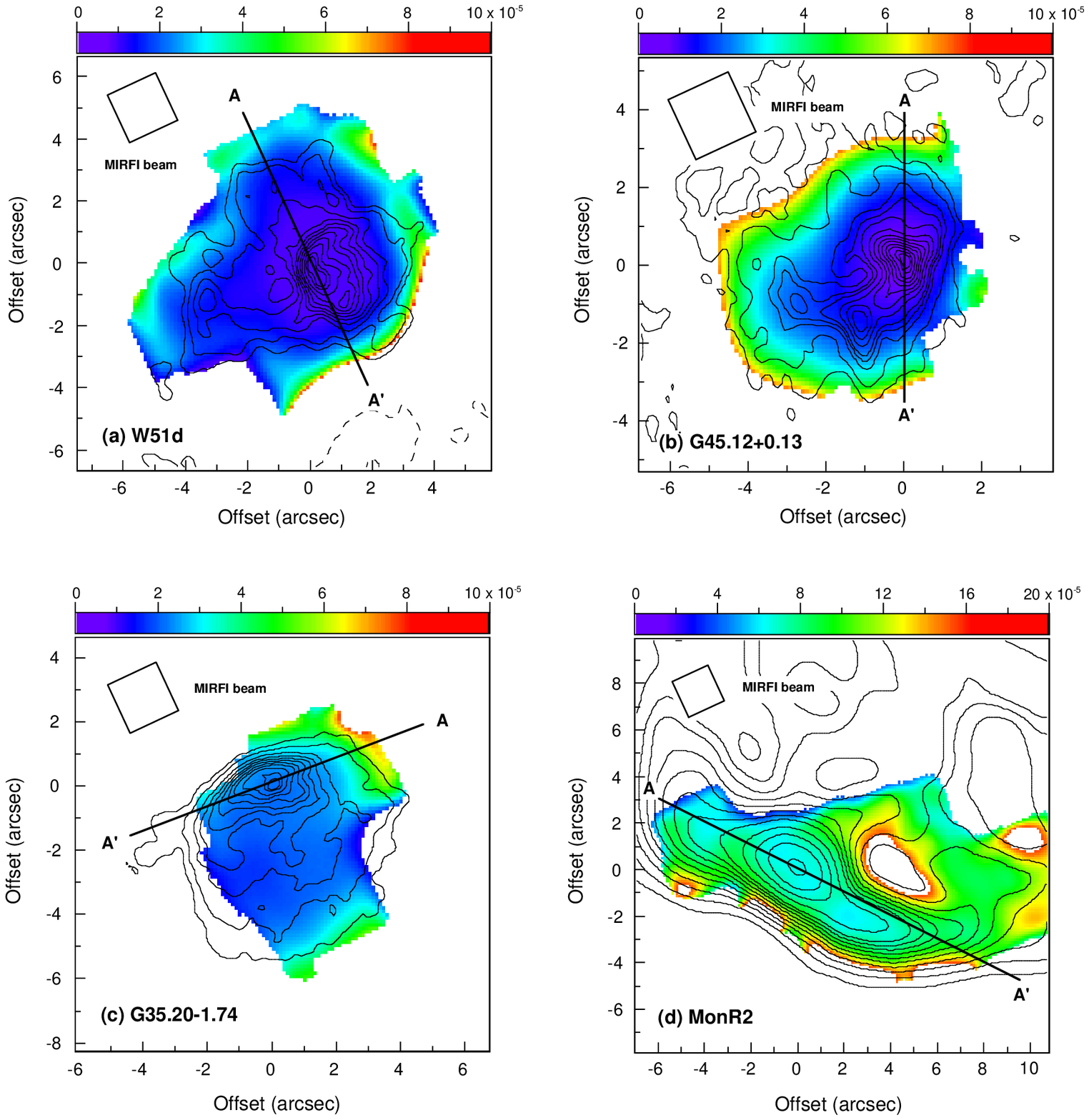]{Ne$^{+}$ abundance ([Ne$^{+}$/H$^{+}$]) maps of four UC H {\small II} regions ($a$) W51d, ($b$) G45.12+0.13, ($c$) G35.20$-$1.74, and ($d$) Mon R2, superposed on the radio contours. The contour intervals are same as those of Figure 2. For Mon R2, the blank spot at west side of the radio peak is caused by the uncertainty of the radio contours. Thick solid lines labeled by A$-$A$^{\prime}$ represent the directions of the cut views shown in Figure 4. \label{fig3}}

\figcaption[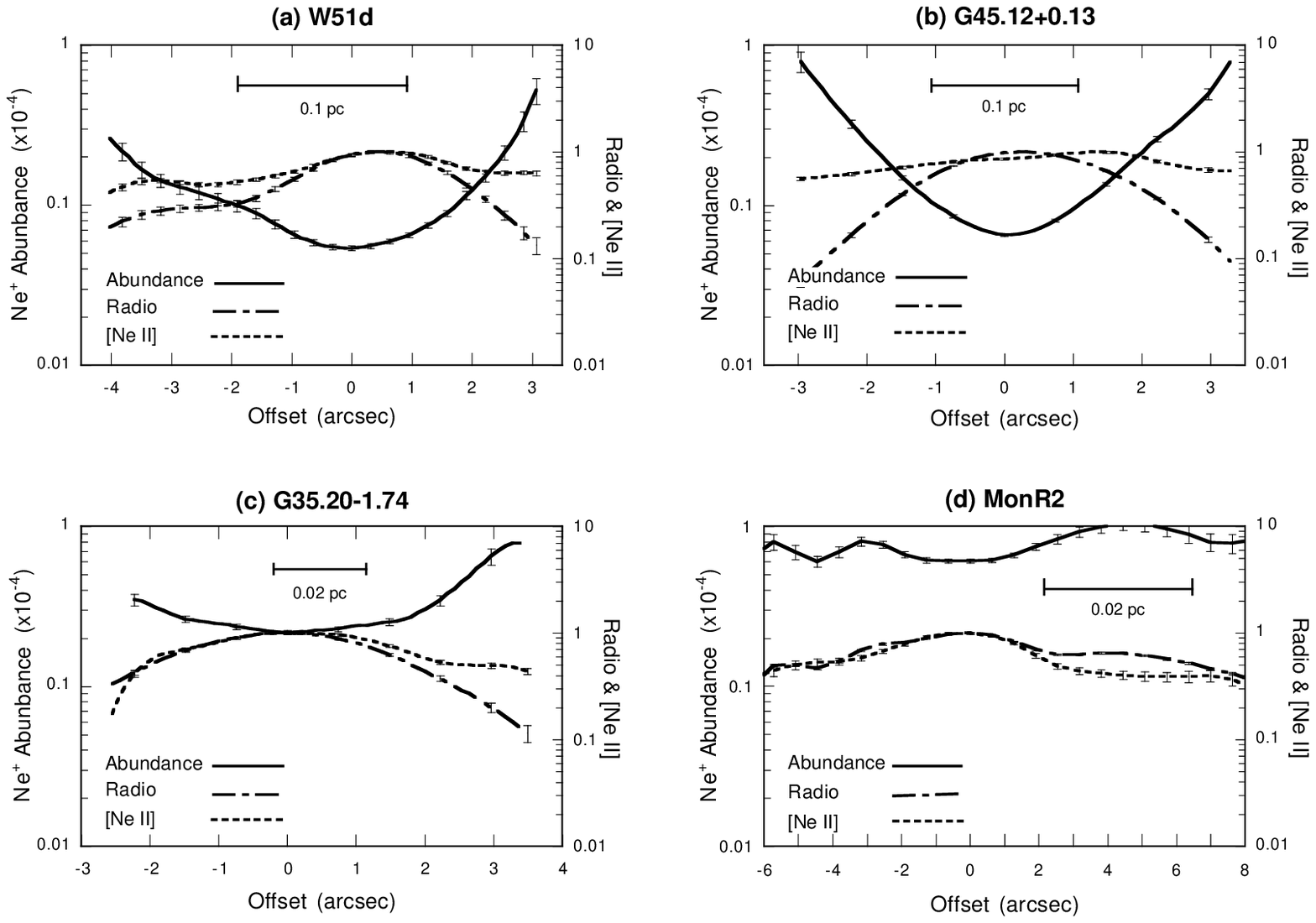]{Cut views of distributions of the [Ne {\small II}] and radio intensities as well as the Ne$^{+}$ abundance. The directions of cuts are shown in Figure 3 (A$-$A$^{\prime}$). The {[Ne {\small II}]} and radio continuum fluxes are normalized by the values at the peak position. Radio profiles are smoothed by $\sim$ 2$\arcsec$ beam ($\sim$ 1 pixel of MIRFI). The error bars represent 1 $\sigma$ statistical errors of both the [Ne {\small II}] and the radio images. \label{fig4}}

\figcaption[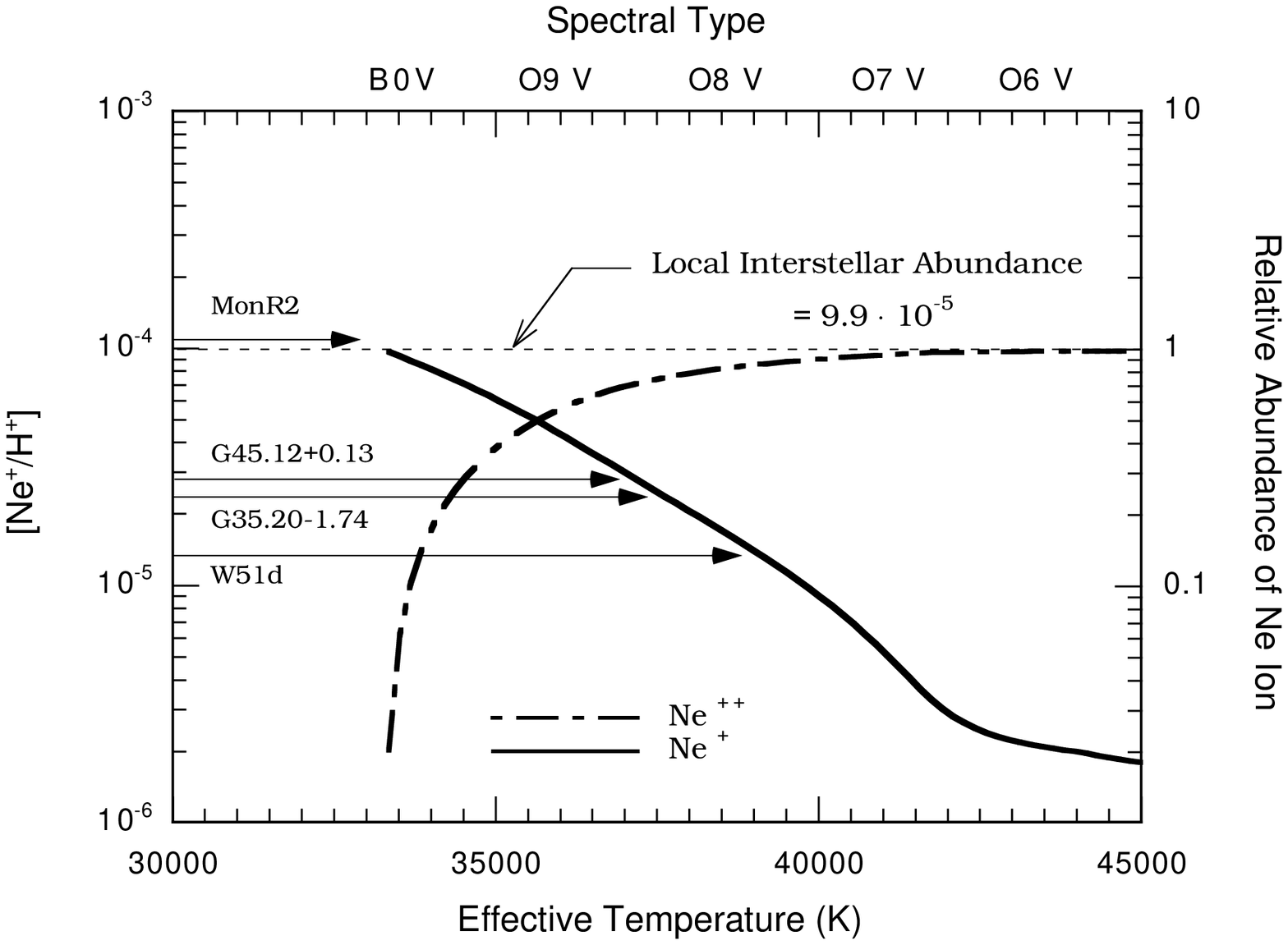]{The Ne$^{+}$ abundance as a function of the effective temperature of the ionizing star calculated from the H {\small II} region model using the {\it CoStar} stellar atmosphere model (Stasi\'nska \& Schaerer 1997). Total abundance of neon (= unity at the right vertical axis) is assumed to be the LIA of neon (= 9.9 $\times$ 10$^{-5}$ at the left vertical axis). The ZAMS spectral types corresponding to the effective temperatures are shown by the top horizontal axis. \label{fig5}}

\figcaption[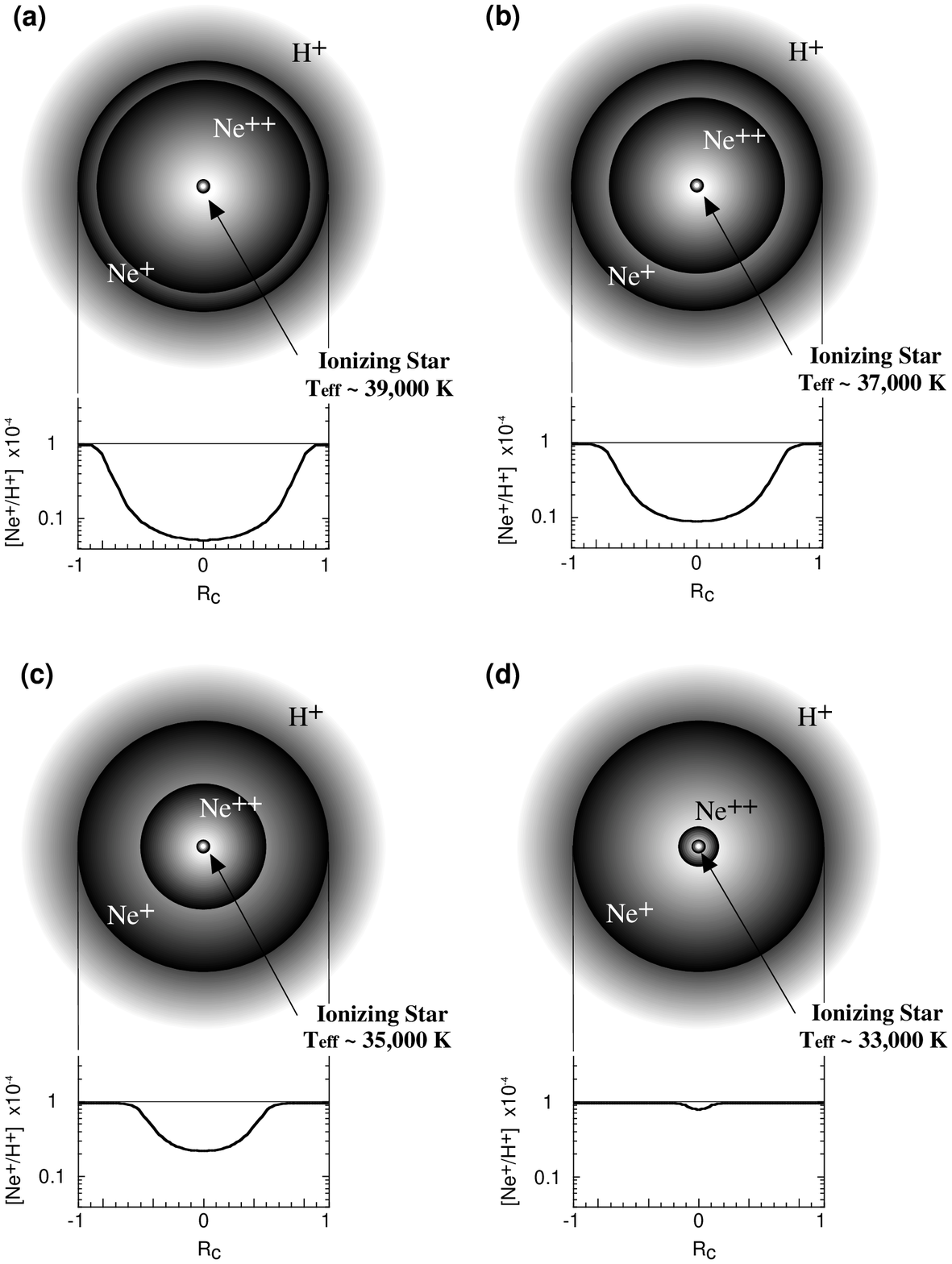]{The schematic model of ionization structure of neon using a simple and spherical H {\small II} region. The effective temperatures of ionizing stars correspond to 39,000 K for ($a$), 37,000 K for ($b$), 35,000 K for ($c$), and 33,000 K for ($d$), respectively. The volume ratio of the Ne$^{++}$ sphere to the Ne$^{+}$ shell is determined from the assumed effective temperature using the model calculation shown in Figure 5. For each panel, the averaged Ne$^{+}$ abundance along the line of sight is also shown. Rc is the distance from the center crossing the equator projected on the sky (normalized by the outer radius). We assume the total abundance of neon is the LIA value. \label{fig6}}

\newpage

\begin{deluxetable}{lccrcc}
\footnotesize
\catcode`?=\active \def?{\phantom{0}}
\tablecolumns{6}
\tablewidth{0pc}
\tablecaption{TARGET LIST \label{tbl-1}}
\tablehead{
\colhead{} & \colhead{{\it IRAS}}  & \colhead{$\alpha$}  & \colhead{$\delta$} & 
\colhead{Distance}  & \colhead{Distance} \\
\colhead{Source} & \colhead{Designation}   & \colhead{(1950.0)}    & \colhead{(1950.0)} & 
\colhead{from the Sun}    & \colhead{to the G.C.} \\
\colhead{} & \colhead{}   & \colhead{(?h??m??s)}    & \colhead{($?\arcdeg ??\arcmin ??\arcsec$)} & 
\colhead{(kpc)}    & \colhead{(kpc)} }
\startdata
W51d \dotfill      & 19213$+$1424\phn & 19?21?22.29 & ? 14?25?15.06 & 7.0? & 6.7  \nl
G45.12$+$0.13 \ldots  & 19111$+$1048\phn & 19?11?06.23 & 10?48?25.80 & 9.7? & 7.1  \nl
G35.20$-$1.74 \ldots & 18592$+$0108\phn & 18?59?14.06 & 01?09?03.19 & 3.2? & 6.2  \nl
Mon R2 \dotfill     & 06053$-$0622\phn & 06?05?19.93 & $-$06?22?39.53 & 0.95 & 9.3  \nl
\enddata
\end{deluxetable}

\begin{deluxetable}{lcccr}
\footnotesize
\catcode`?=\active \def?{\phantom{0}}
\tablecolumns{5}
\tablewidth{0pc}
\tablecaption{OBSERVATION LOG \label{tbl-2}}
\tablehead{
\colhead{Source} & \colhead{Date} & \colhead{Frames\tablenotemark{\ast}} & 
\colhead{Airmass}  & \colhead{Guide Star?}} 
\startdata
W51d \dotfill   & Oct. ?9 ?1997 & 6 &  1.13 $-$ 1.89 & 31 Aql.?? \nl
G45.12$+$0.13 \ldots\ldots & Oct. ?5 ?1997 &8 & 1.16 $-$ 1.20 & $\gamma$ Aql.?? \nl
G35.20$-$1.74 \ldots\ldots & Oct. ?5 ?1997& 3 & 1.49 $-$ 1.58  & tt Aql.?? \nl
\  & Oct. ?9 ?1997& 2 &  1.31 $-$ 1.37 & \nodata???  \nl
Mon R2 \dotfill  & Oct. ?4 ?1997& 3 &  1.48 $-$ 1.49 & SAO132884  \nl
\  & Oct. ?8 ?1997& 4 &  1.48 $-$ 1.58 & \nodata???  \nl
\  & Oct. ?9 ?1997& 8 &  1.48 $-$ 1.78 & \nodata???  \nl
\  & Oct. 10 ?1997& 6 &  1.48 $-$ 1.67 & \nodata???  \nl
\enddata
\tablenotetext{\ast}{?See text for the definition.}
\end{deluxetable}

\begin{deluxetable}{lccrccccccc}
\scriptsize
\tablenum{3}
\catcode`?=\active \def?{\phantom{0}}
\tablewidth{0pt}
\tablecaption{DERIVED PHYSICAL PROPERTIES \label{tbl-3}}
\tablehead{
\colhead{} & \colhead{} & \multicolumn{2}{c}{F[Ne II] flux}  & \colhead{} & \multicolumn{5}{c}{Radio parameters}  &\colhead{Abundance}\\[+4pt]
\cline{3-4} \cline{6-10} \\[-6pt]
\colhead{Source}   & \colhead{Integ. Area} & \colhead{observed} & \colhead{corrected\tablenotemark{?a}} & \colhead{} & \colhead{Radio flux\tablenotemark{?b}} & \colhead{T$_{b}$} & & \colhead{$\tau$} & \colhead{EM}  & \colhead{{[Ne$^{+}$/H$^{+}$]}}\\[+2pt]
\colhead{}    & \colhead{(10$^{-11}$ sr)} & \multicolumn{2}{c}{(10$^{-18}$ W cm$^{-2})$} & \colhead{} & \colhead{(mJy)} & \colhead{(K)} & & \colhead{} & \colhead{(10$^{7}$ pc cm$^{-6}$)}  & \colhead{(10$^{-5}$)}\\[+6pt]
\cline{1-11} \\[-9pt]
\multicolumn{4}{r}{\bf TOTAL} & & \multicolumn{5}{l}{\bf REGION\tablenotemark{?c}}\vspace*{-3.8mm}\\}
\startdata
W51d $\dotfill$   & 164 & 27.9\ $\pm$\ 6.6 & 55.3\ $\pm$\ 13.0 & & 13900\ $\pm$\ 1400 & 1240\ $\pm$\ 120 & & 0.18\ $\pm$\ 0.02 & 11.4\ $\pm$\ 1.1 & ?1.3 $\pm$\ 0.3 \nl
G45.12$+$0.13 $\dotfill$ &155 & 18.2\ $\pm$\ 4.2 & 42.7\ $\pm$\ 9.8? & & ?5210\ $\pm$\ 520? & ?490\ $\pm$\ 50? & &  0.06\ $\pm$\ 0.01 & ?4.3\ $\pm$\ 0.4 & ?2.8 $\pm$\ 0.7 \nl
G35.20$-$1.74 $\dotfill$ & ?95 & ?9.5\ $\pm$\ 2.3 & 20.9\ $\pm$\ 5.1? & & ?3140\ $\pm$\ 310? & ?480\ $\pm$\ 50? & &  0.07\ $\pm$\ 0.01 & ?4.1\ $\pm$\ 0.4 & ?2.3 $\pm$\ 0.6 \nl
Mon R2 $\dotfill$  & 257 & 20.3\ $\pm$\ 4.7 & 44.5\ $\pm$\ 10.3 & & ?1610\ $\pm$\ 160? &  ?870\ $\pm$\ 90? & &  0.12\ $\pm$\ 0.01 & ?0.7\ $\pm$\ 0.1 & 10.7 $\pm$\ 2.7 \nl
\vspace*{-0.7mm}\\
[-6pt]\hline \vspace*{-3.2mm}\\
 & & & {\bf PEAK} & & {\bf REGION\tablenotemark{?c}\ \ } & & & & & \nl
\vspace*{0.6mm}\\
[-10pt]\hline \vspace*{-2.6mm}\\
W51d $\dotfill$   & 6.8 & 2.4\ $\pm$\ 0.6 & 4.7\ $\pm$\ 1.1 & & 2340\ $\pm$\ 210 & 5040\ $\pm$\ 470 & &  1.06\ $\pm$\ 0.18 & 66.8\ $\pm$\ 11.2 & 0.45 $\pm$\ 0.13 \nl
G45.12$+$0.13 A \ldots & 6.8 & 1.2\ $\pm$\ 0.3 & 2.7\ $\pm$\ 0.6 & & 1050\ $\pm$\ 130 & 2250\ $\pm$\ 270 &  & 0.33\ $\pm$\ 0.05 & 22.5\ $\pm$\ 3.2? & 0.79 $\pm$\ 0.21 \nl
????????????B \ldots & 6.8 & 1.5\ $\pm$\ 0.3 & 3.5\ $\pm$\ 0.8 & & ?750\ $\pm$\ 50? & 1610\ $\pm$\ 120 & & 0.23\ $\pm$\ 0.02 & 15.3\ $\pm$\ 1.3? & 1.5? $\pm$\ 0.4? \nl
G35.20$-$1.74 $\dotfill$ & 6.8 & 1.4\ $\pm$\ 0.3 & 3.1\ $\pm$\ 0.8 & & ?520\ $\pm$\ 60? & 1130\ $\pm$\ 130 & &  0.16\ $\pm$\ 0.02 & 10.2\ $\pm$\ 1.3? & 2.0? $\pm$\ 0.5? \nl
Mon R2 $\dotfill$  & 6.8 & 1.0\ $\pm$\ 0.2 & 2.2\ $\pm$\ 0.5 & & ?100\ $\pm$\ 2?? &  2060\ $\pm$\ 40? & &  0.32\ $\pm$\ 0.01 & ?1.9\ $\pm$\ 0.04 & 7.7? $\pm$\ 1.8? \\
\enddata
\tablecomments{?We use the electron temperature as follows; W51d : 7,700K (Mehringer 1994), G45.12+0.13 : 8,000K (Lumsden \& Puxley 1996), G35.20$-$1.74 : 7,500 K (Wink, Wilson, \& Bieging 1983), Mon R2 : 7,600 K (Dowens et al. 1975).}
\tablenotetext{a}{?Corrected flux for interstellar extinction.}
\tablenotetext{b}{?Radio continuum fluxes at 15 GHz for all sources except for Mon R2. As for Mon R2, 5 GHz flux is adopted (Wood \& Churchwell 1989).}
\tablenotetext{c}{?TOTAL REGIONs are defined as the regions where the [Ne II] flux is appreciably detected (above 3$\sigma$). PEAK REGIONs correspond to the MIRFI 1 pixel regions (1$\/\farcs\/$7 $\times$ 1$\/\farcs\/$7) at the radio peaks except for G45.12+0.13 B, which is the [Ne II] emission peak.}
\end{deluxetable}

\begin{deluxetable}{clllll}
\footnotesize
\tablenum{4}
\catcode`?=\active \def?{\phantom{0}}
\tablewidth{0pt}
\tablecaption{SPECTRAL TYPE OF IONIZING STAR \label{tbl-4}}
\tablehead{
\colhead{Source} & \multicolumn{2}{c}{FIR Luminosity \tablenotemark{a}} & \colhead{Radio Luminosity \tablenotemark{b}} & \colhead{MIR Line Ratio \tablenotemark{c}} & \colhead{This Work} \\
\colhead{} & \colhead{Single Star} & \colhead{Cluster}  & \colhead{Single Star} & \colhead{}  & \colhead{}}
\startdata
W51d\dotfill  & ???O4 & ?O4.5 & ??????O5.5  & & ?{\ \ \ \ }O7.5V$-$O8V \nl
G45.12+0.13\ldots  & ???O4 & ?O5  & ??????O6.5  & ????O9.5V & ?{\ \ \ \ }O8V$-$O8.5V \nl
G35.20$-$1.74\ldots  & ???O6 & ?O9.5  & ??????O9 & & ?{\ \ \ \ }O8V$-$O9V \nl
Mon R2\dotfill & & & ??????B0 \tablenotemark{d}?? & ????B0V & ??$\leq$B0V \tablenotemark{e} \nl
\enddata
\tablenotetext{a}{ {\it IRAS\/} PSC (Wood \& Churchwell 1989).}
\tablenotetext{b}{ Converted from $N_{\rm Lyc}$ luminosity (Wood \& Churchwell 1989).}
\tablenotetext{c}{ Converted from {[Ne$^{+}$/Ne$^{++}$]} given by Simpson \& Rubin 1990 using the calculations based on the {\it CoStar} stellar atmosphere model (Stasi\'nska \& Schaerer 1997).}
\tablenotetext{d}{ Downes et al. 1975.}
\tablenotetext{e}{ The {\it CoStar} model grids for T$_{\rm eff}$ $<$ 33,000K ($<$ B0) are not available.}
\end{deluxetable}

\end{document}